# Safeguarding People's Financial Health in Metaverse with Emotionally Intelligent Virtual Buddy

SYED ALI ASIF, EMMA CAO, HANG CHEN, and CHIEN-CHUNG SHEN, University of Delaware, USA

YAN-MING CHIOU, SRI, USA

**ACM Reference Format:**
Syed Ali Asif, Emma Cao, Hang Chen, Chien-Chung Shen, and Yan-Ming Chiou. 2024. Safeguarding People's Financial Health in Metaverse with Emotionally Intelligent Virtual Buddy. In *CHI Workshop on Novel Approaches for Understanding and Mitigating Emerging New Harms in Immersive and Embodied Virtual Spaces, May 11, 2024, Honolulu, Hawaii, USA.* ACM, New York, NY, USA, 2 pages. https://doi.org/XXXXXXX.XXXXXXX

## POSITION STATEMENT

The Metaverse, an immersive virtual world, has emerged as a shared space where people engage in various activities ranging from social interactions to commerce. Cryptocurrencies [3] and Non-Fungible Tokens (NFTs) [6] play pivotal roles within this virtual realm, reshaping interactions and transactions. Cryptocurrencies, utilizing cryptographic techniques for security, enable decentralized and secure transactions, and NFTs represent ownership or proof of authenticity of unique digital assets through the blockchain technology. While NFTs and cryptocurrencies offer innovative opportunities for ownership, trading, and monetization within the metaverse, their use also introduces potential risks and negative consequences, such as financial scams and fraud, highlighting the need for users to exercise caution and diligence in their virtual transactions.

As the adoption of cryptocurrencies and NFTs proliferates within the metaverse, so too do the opportunities for exploitation by malicious actors [7]. The decentralized and pseudonymous nature of these technologies can obscure the identities of scammers, making it challenging for victims to seek recourse. Moreover, the absence of regulatory oversight often exacerbates the risks associated with fraudulent activities. Therefore, users must remain vigilant, conduct thorough research, and verify the authenticity of transactions before engaging in any virtual exchange. By adopting a proactive approach to security and staying informed about emerging threats, individuals can better protect themselves against the ever-present specter of scams and fraud involving cryptocurrency and NFT. The following example NFT scam illustrates how financial health can be compromised within the metaverse.

> Sarah, an enthusiastic collector of NFTs in the metaverse, becomes a target of an NFT scam when she engages with an avatar who poses as the renowned digital artist Beeple. Intrigued by the avatar's claim, she is enticed into a private room where she is shown an impressive digital collection. Falling for the deception, she purchases a rare piece of digital art at a meager price, approximately half a Bitcoin. Filled with excitement over her purchase, Sarah's spirits were quickly dashed when her friend raised concerns about the authenticity of her NFT. Upon further investigation, she realized she had been scammed and bought a fake. As a consequence, she experienced financial difficulties for a considerable period.

To prevent scenarios like Sarah's in the metaverse, new technologies are needed to raise awareness and safeguard users from potential harms. Sentiment analysis and Generative AI offer promising solutions. Sentiment analysis [2] involves the process of identifying and categorizing the attitude toward a situation or event expressed in user-generated content, providing valuable insights into their underlying thinking and feeling, and has also proven effective in fraud detection [4, 5]. On the other hand, Generative AI [1] is capable of generating new content based on patterns and data inputs, enabling prompt and effective responses tailored to individual user needs. These technologies can analyze interactions, detect suspicious activities, and alert users about potential risks and negative consequences. Integrating them can proactively mitigate scams and fraud, fostering a safer virtual environment.

In our endeavor to advance people's financial resilience in the metaverse against the risks posed by cryptocurrency and NFT fraud and scams, we have developed an empathetic avatar, referred to as a virtual buddy. This avatar leverages sentiment analysis and Generative AI to establish rapport and facilitate social interactions within the metaverse.

Initial post-surveys conducted with a limited sample size have shown promising results regarding the perceived helpfulness of the application-agnostic virtual buddy. Our research now aims to rigorously examine the avatar's potential to mitigate financial risks based on preliminary evidence. In addition, we plan to employ neurophysiological approaches, including EEG measurements, in conjunction with pre- and post-surveys to obtain quantitative observations. Additionally, qualitative observations will be conducted during metaverse sessions to supplement the quantitative data with insights into user behavior and interactions with the virtual buddy.

By adopting this comprehensive methodology, we aim to provide a nuanced understanding of the virtual buddy's efficacy in safeguarding users against cryptocurrency and NFT scams within the metaverse. This research contributes to the advancement of virtual companion technology and holds implications for developing strategies to enhance financial literacy and security within the emerging digital realm.